\documentstyle[12pt,psfig]{article}

\begin{document}

 \title{How Can Computer Simulations Contribute
                      to Understand the Static Structure of Glasses?
\footnote{To appear in {\it Analysis of Composition and Structure of
Glass, and Glass Ceramics} Eds.: H. Bach and D. Krause (Springer,
Berlin, 1998)}}

 \vspace{2cm}

  \author{Kurt Binder and Walter Kob\\
            Institut f\"ur Physik,
            Jo\-han\-nes-Gu\-ten\-berg-Uni\-ver\-si\-t\"at Mainz   \\
            D-55099 Mainz, Staudinger Weg 7, Germany}

 \date{November 30, 1997}

  \maketitle

 \subsubsection*{Introduction: The Molecular Dynamics Method}
Many different computational techniques are denoted as ``computer
simulation'', from ``macroscopic'' techniques, such as finite element
methods to calculate mechanical properties, to ``microscopic''
techniques that deal with nuclei and electrons to calculate the
electronic structure of materials (e.g., the density functional method
[1]). However, in the present article we mean by {\it structure} the
geometric arrangement of atoms considering typical length scales from
0.2~\AA~to 20~\AA. In particular, we are interested in how this
geometric structure depends on the chemical constituents (SiO$_2$,
B$_2$O$_3$, $\ldots$), thermodynamic variables (temperature $T$,
pressure $p$), cooling history by which the glass was prepared, etc. As
input to such a simulation on the atomistic scale, one wishes to take a
suitable {\it force field} that describes the interaction between the
atoms, treating them like {\it point particles in classical mechanics}.
Of course, this neglect of all quantum effects is a severe limitation
of the approach that we are going to describe, and in principle it can
be avoided by more accurate methods: the Car-Parrinello method [2]
combines molecular dynamics with the density functional treatment of
electronic structure which depends properly on the coordinates of the
nuclei, and thus avoids the approximations inherent in the force field
to some extent --- the path integral Monte Carlo method [3] takes
into account the Heisenberg uncertainty principle (i.e., the fact that
one cannot specify precisely simultaneously positions and momenta of
the nuclei, since the latter are spread out over a region of the order
of the de Broglie wavelength), and thus allows a correct treatment of
the thermal properties of solids at low temperatures.

Although the application of either technique to materials like SiO$_2$
may yield promising results in the near future, we confine ourselves
here to the molecular dynamics (MD) methods in the framework of
classical statistical mechanics [4]. Unlike the techniques proposed
more recently [2,3], applications of MD to simulate glasses have
been developed since a long time [5], and we feel that recent
applications [6,7] allow a rather precise and detailed description
of both the structure and the thermal properties of glasses.

Basically, the MD simulation tries to mimic the experimental procedure
by which a glass is produced from the melt that is cooled down
sufficiently fast to bypass crystallization. However, as will be
discussed below, an important difference between experiments and
simulations is that in the MD method cooling rates are many
orders of magnitude faster than in reality, and these extremely fast
cooling schedules do have an effect on the physical properties of the
``glass'' observable in such a ``computer experiment'' [6]. To bypass
this difficulty, there are actually many attempts to directly create
the amorphous structure of silica glass or other so-called ``continuous
random networks'' by direct construction [8]. For example, one uses
topological rules (e.g. in amorphous SiO$_2$ each silicon atom has to
have four oxygen atoms as neighbors, each oxygen atom has to have two
silicon atoms as neighbors, the length of the covalent Si--O-bond is
taken as constant and used as an input parameter) but allows for some
disorder (e.g. the angles between covalent bonds are taken from some
ad-hoc chosen distributions which are often optimized to match the
experimental radial pair distribution function [8]). We feel,
however, that despite its successes to reproduce certain known facts
about the geometrical structure of glasses this heuristic approach
suffers from unknown (and hence uncontrolled) errors, unlike MD which
--- in the framework of classical statistical mechanics --- is a
``first principles''-method.  We also mention that an alternative
``first principles''-method is Monte Carlo sampling [9]. There one
generates configurations of the model system distributed according to
the canonical distribution function of statistical mechanics, i.e.
proportional to $\exp(- {\cal{H}}/k_BT)$ where ${\cal{H}}$ is the
Hamiltonian and $k_B$ is the Boltzmann's constant. While this method is
widely used, particularly for suitably coarse grained models, where
many of the atomistic degrees of freedom are disregarded, it does not
seem to offer any particular advantage for the atomistic simulation of
inorganic glasses. Hence this technique will not be described here
either.

\subsubsection*{Basic Features of a Molecular Dynamics Program: an
Introduction for the Novice}
The equations of motions that are solved in elementary MD are Newton's
equations for $N$ atoms with Cartesian coordinates $\{\vec{r}_i\}$,
$i=1$, $\ldots$, $N$, $\nabla_i$ being the gradient operator
with respect to $\vec{r}_i$,
\begin{equation}   
m_i \ddot{\vec{r}}_i =
- \nabla_i U \equiv {\vec{f}_i}, \quad
\vec{v}_i \equiv \dot{\vec{r}}_i
= d \vec{r}_i/dt, \qquad \qquad \qquad
\end{equation}
where $m_i$ is the mass of particle $i$ and $\vec{f}_i$ is the force on
atom $i$, derived from the potential energy function $U$ of the
system,
\begin{equation}  
U(\vec{r}_1, \vec{r}_2, \ldots, \vec{r}_N) = \sum^{N}_{i=1} 
\sum^{N}_{j>i} u (|\vec{r}_i - \vec{r}_j|).
\end{equation}
In Eq.(2) we have made the approximation that $U$ is simply
the sum of {\it pair} potentials between atoms, which clearly will not
be true in general, but which greatly simplifies the calculations since
then the force ${\vec{f}_i}$ simply becomes
\begin{equation}   
\vec{f}_i = - \sum^{N}_{j(\neq i)=1} \, \partial u
(|\vec{r}_i- \vec{r}_j|)/\partial \vec{r}_i =
\sum^{N}_{j (\neq i)=1} \, \vec{f}_{ij}.
\end{equation}
In fact, it is  not at all obvious that the directional covalent bonds
in materials such as SiO$_2$, B$_2$O$_3$ etc. can be reliably modelled
by Eq.(2), and indeed potentials are sometimes used that explicitly
include three-body terms that keep a bond angle $\theta$ close to its
equilibrium value $\theta_0$, e.g. [10, 11]
\begin{equation}   
u_{\theta} = \frac{1}{2} K_{\theta} (\theta - \theta_0)^2.
\end{equation}
However, while in the case of B$_2$O$_3$ even inclusion of terms such
as Eq.(4) still does not allow a fully satisfactory modelling of
glassy structures [11], in the case of SiO$_2$ the situation seems to
be fortunately much simpler, and a pairwise potential proposed by {\it
van Beest et al.} [12], henceforth referred to as ``BKS potential'',
already gives reliable results [6,13,14]. With the abbreviation
$r_{ij} = |\vec{r}_i -\vec{r}_j|$, the BKS potential is given by
\begin{equation}   
u(r_{ij}) =  q_i q_j e^2/r_{ij} +
A_{ij} \exp (-B_{ij} r_{ij}) - C_{ij} r_{ij}^{-6} \quad
{\mbox{with}}\quad i,j \in \{{\mbox{Si,O}}\}. 
\end{equation} 
Here $e$ is the charge of an electron. Note that $q_{{\rm O}}$ and
$q_{{\rm Si}}$ are {\it effective charges}, i.e. are given by $-1.2$
and 2.4, respectively ($q_{{\rm Si}}+2 q_{{\rm O}} = 0$ is required
because of charge neutrality, of course). Apart from this (pseudo-)
Coulomb interaction between every pairs of atoms, the only short range
interactions present are the ones between Si--O pairs and O--O pairs,
with parameters $A_{{\rm OO}} \approx 1389$~eV, $A_{{\rm SiO}} \approx
18004$~eV, $B_{{\rm OO}} \approx 2.76$~${\mbox{\AA}}^{-1}$, $B_{{\rm
SiO}} \approx 4.873$~${\mbox{\AA}}^{-1}$, and $C_{{\rm OO}} =
175$~eV${\mbox{\AA}}^{-6}$, $C_{{\rm SiO}} \approx
133.6$~eV${\mbox{\AA}}^{-6}$~[3.12].

As is well-known, Newton's equations conserve the total energy $E=U+K$,
where $K$ is the kinetic energy,
\begin{equation}  
K = \sum^{N}_{i=1} \frac{1}{2} m_i \left(\dot{\vec{r}}_i\right)^2.
\end{equation}
If one solves these equations numerically for a (closed) system of
$N$ atoms in a box of volume $V$, one generates a trajectory through
phase space from which one can compute the time average of various
physical observables.  Since $E$, $N$, $V$ are fixed one therefore 
realizes the
microcanonical ensemble of statistical mechanics. Since we deal with
classical statistical mechanics, temperature $T$ can be inferred from
the equipartition theorem and thus, if the system is in thermal
equilibrium throughout the MD ``run'', we have
\begin{equation}  
{\overline{K}} = 3 N k_BT/2.
\end{equation}
(By a bar we henceforth denote the time average along the trajectory.)
Similarly, pressure can be computed from the virial theorem
\begin{equation}   
p = \frac{1}{V}\left\{ N  k_BT + (1/3) \sum^{N}_{i=1}
{\overline{{\vec{r}_i} \cdot {\vec{f}_i}}} \right\} .
\end{equation}

In statistical mechanics, one often wishes to use an ensemble where $V$
and $T$ or $p$ and $T$ rather than $V$ and $E$ are the given
independent variables.  While in the thermodynamic limit these various
ensembles yield equivalent results, for finite $N$ systematic
differences occur.  Furthermore it is also important to note that any
time average ${\overline{A}}$ can be estimated only to within a certain
``statistical error'', because of the finite time span covered by the
MD run. Thus generalizations of MD for the microcanonical ensemble to
these other ensembles are useful and widely used (in ``constant
pressure'' MD one needs to couple the system to a piston, in ``constant
temperature'' MD to a thermostat, etc. [4]).  For derivations and
details about these extensions, we refer to the literature [4]. Here
we only mention another aspect related to the finite size of the
simulated system, and this is the aspect of boundary conditions on the
surface of the simulation box (which for fluids and amorphous systems
usually is taken of cubic shape, $V=L^3$, where $L$ is the linear
dimension of the box). Since one is typically interested in bulk
behavior of the simulated material, and one wishes to work with only
about $10^3$ to $10^4$ atoms due to limited computer resources, it is
crucial to make disturbances due to these boundaries as small as
possible.  This is achieved by using periodic boundary condition, i.e.
the box is surrounded by identical images of itself centered at all
integer multiples of $L$ (when the center of the original box is taken
as the origin of the coordinate system) and thus the system becomes
quasi-infinite. Of course, only the particles in the central box are
independent and need to be handled by Newton's equations, but if a
particle leaves the box through, say, the surface on its right, the
next image of that particle enters the box from the left. While these
surfaces thus have no physical effects whatsoever, i.e. surface effects
are completely eliminated, it is important to realize that effects
associated with the finiteness of $N$ are still present. With respect
to thermodynamic properties, such finite size effects are well known
(e.g.  ``rounding'' of phase transitions [9]), but one must also be
aware that the dynamics of the system may be affected as well (e.g. a
phonon mode traveling to the right may re-enter the box from the left
and interfere with itself, etc.  [15]).

The periodic boundary condition must also be correctly treated when one
considers interactions among distant particles. For slowly decaying
potentials, such as the Coulomb part in Eq.(5), it is essential to
sum the interactions not only over all particles in the primary box
but also over all the periodic images --- every atom interacts thus
with an infinite number of images of all the other particles!
This seemingly difficult task, however, is efficiently solved by the
Ewald method [4]. For the short range part, however, one may even
follow the simpler strategy to cut the potential off if the distance
$r$ exceeds a maximum distance $r_{\mbox{\scriptsize max}}$ (e.g.,
for the O--O-interaction in [6] $r_{\mbox{\scriptsize max}}= 5.5$ {\AA}
could be used since then the strength of the potential has decreased
to about 1 \% of its value at the preferred O--O distance).

As a last point, we emphasize that integration of Newton's equations,
Eq.(1), for a many-particle system of interest cannot be done
exactly, but only approximately, due to the finiteness of the time step
$\delta t$ of the numerical integration procedure. By making $\delta t$
sufficiently small this error can be made as small as desired --- but
then it becomes increasingly difficult to propagate in time up to time
scales of physical interest. Thus it is important to choose a ``good''
integration algorithm. Several suitable algorithms exist, but here only
the standard Verlet algorithm is quoted [4], which is correct up to
order $(\delta t)^3$,
\begin{eqnarray}   
\vec{r}_i(t+ \delta t) & = & \vec{r}_i (t) + \delta t \vec{v}_i (t)
+ \frac{1}{2m_i} (\delta t)^2 \vec{f}_i (t),      \\
\vec{v}_i (t + \delta t) & = & \vec{v}_i(t) + \frac{\delta t}{2 m_i}
[\vec{f}_i(t) + \vec{f}_i (t + \delta t)].
\end{eqnarray}
Note that the propagation of the velocities, Eq.(10), requires that the
forces are known both at the present time $t$ and at the next time step
$t + \delta t$, which can of course be calculated from $\vec{r}_i(t +
\delta t)$ using Eq.(3). This algorithm is time reversible and
conserves the phase space volume of a continuous set of phase space
points, as is required for time averaging in statistical mechanics
[4].  Nevertheless, it is clearly necessary that $\delta t$ is at
least about two orders of magnitude smaller than the time constant of
elementary  physical  motions in the system, e.g. vibration times of
covalent Si--O-bonds or bond angles, which are significantly less than
one pico second:  Therefore it becomes necessary to choose a time-step
$\delta t$ as small as $\delta t = 1.6 \times 10^{-15}$ sec to get
reliable results! Thus even very long MD runs (several millions of time
steps) can span only a very short interval of physical time, of the
order of $10$~nano sec. This is not only true for SiO$_2$ but for MD
simulations of atomistic models of real materials in general. While
experimental techniques such as neutron scattering [16] are
constrained to a similarly restricted ``observation time window'', the
important distinction is that in MD the same short time scale is
available not only for observation but also for the {\it equilibration
of the structure}. Therefore the way how a structure for an amorphous
material studied by MD was prepared needs careful attention ---
although this is an obvious point, it often is ``swept under the rug''
in the literature!

For example in the results described in the next section, this
``preparation of the system'' was done by making first an MD run of
40,000 time steps at an initial temperature $T_i= 7000 K$. This length
is more than enough to equilibrate the system at this extremely high
temperature. Subsequently the system was cooled, at zero pressure, by
coupling it to a heat bath whose temperature $T_b(t)$ was decreased
linearly in time, $T_b(t) = T_i - \gamma t$, with cooling rates
$\gamma$ from $1.14 \times 10^{15}$~K/s to $4.44 \times 10^{12}$~K/s.
Even the slowest cooling rate is many orders of magnitude larger than
the ones used in the laboratory. However, it is currently not possible
to simulate a quench of the system with cooling rates that are
significantly smaller than the ones used here, since for the smallest
cooling rate one simulation run of the system (which contained 334 Si-
and 668 O-atoms in the central box) needs already about 340 h of CPU
time on an IBM~ RS6000/370 workstation, and one needs to take a sample
average over at least 10 independent runs (starting from different
initial configurations) to improve the statistical accuracy of the
``simulation data''.

\subsubsection*{A Case Study: Cooling-Rate Dependence of the Structure of
amorphous SiO$_2$}
Here we briefly review some typical results that were recently obtained
by {\it Vollmayr et al.} [6]. As an example for macroscopic
properties we show in Fig.~1 the temperature dependence of the
density for the different cooling rates studied. One sees that, at very
high temperatures, all curves (apart from the ones for the three
fastest cooling rates) fall onto a master curve, the equilibrium curve.
However, for the fastest cooling rates the system falls out of
equilibrium during the cooling process immediately, demonstrating that
even at extremely high temperatures such runs do not give any reliable
information on the thermal equilibrium behavior of the system.

The curves corresponding to the smallest cooling rates show a maximum
at around 4800~K, i.e. a density anomaly. Qualitatively, but not
quantitatively, this anomaly is in agreement with experiment (the
experimental value for the temperature of the density maximum is
considerably lower, namely 1820~K [17]). It is unlikely that this
quantitative discrepancy is related to the too fast cooling since the
location of the maximum is independent of the cooling rate, if the
latter is sufficiently small. More probably the discrepancy reflects an
inaccuracy of the potential. However, for different potentials this
density anomaly is present only at even higher temperatures or not
present at all [18]: thus nothing would be gained by using instead of
Eq.(5) one of these (typically more complicated) alternative
potentials.

For intermediate and small values of $\gamma$, the density
decreases upon cooling (after having passed the maximum) and reaches a
minimum. At even lower temperatures the curves become, within the
accuracy of our data, straight lines with negative slope. The value of
this (positive) thermal volume expansion coefficient $\alpha$ decreases
with decreasing cooling rate from about $\alpha_p \approx 7 \times
10^{-6}$~K$^{-1}$ for $\gamma = 1.14 \times 10^{15}$~K/s to $\alpha_p
\approx 2.5 \times 10^{-6}$~K$^{-1}$ for $\gamma = 4.44 \times
10^{15}$~K/s. Experimental values at and above room temperature are
[17] $\alpha_p \approx 0.55 \times 10^{-6}$~K$^{-1}$.  Note, however,
that quantum statistical mechanics requires that $\alpha_p (T \to 0)
\to 0$, i.e. the density must reach its groundstate value with
vanishing slope. This feature can never be reproduced by a purely
classical calculation such as MD. In fact deviations from classical
statistical mechanics are expected at temperatures of the order of the
Debye temperature, which is roughly $T_{\Theta} \approx 1200$~K in
SiO$_2$ [19]. A similar caveat also applies to the specific heat
$C_p$, which reaches in the MD simulation [6] at low temperatures a
value of about 1.25~J/gK, close to the value of the classical
Dulong-Petit law (which would yield 1.236~J/gK): classical MD (as well
as classical Monte Carlo) can never reproduce the vanishing of the
specific heat as $T \to 0$. (In principle, this can be achieved by the
quantum-mechanical path integral Monte Carlo method [3] but this
approach has so far only been applied to crystals and not yet to
glasses.)

While experimentally [19] $C_p$ rises rather sharply at the glass
transition temperature from about 1.23~J/gK to about 1.5~J/gK, the
simulation yields a smooth increase at about $T=3000$~K to a maximum
value of $C_p^{\mbox{\scriptsize max}} \approx 2$~J/gK near $T=4000$~K;
the latter behavior is thus not backed by experiment and may partially
be due to artifacts of the too rapid cooling and partially due to
inadequacies of the potential~[5].

Although the MD simulation of molten SiO$_2$ using the BKS potential
gives neither at very high temperatures nor at low temperatures a {\it
quantitatively} accurate description of thermal properties (density,
specific heat, etc.), it nevertheless gives a very good description of
the geometric structure of the glass. This is evident if one examines
the radial distribution functions $g_{\alpha \beta} (r)$ between
species $\alpha$ and $\beta$, Fig. 2 (see Ref. [20] for theoretical
background on this quantity). One recognizes that with decreasing
cooling rate the structural order at short and intermediate distances
(i.e. $r \leq 8$ {\AA}) increases, since the peaks and minima of
$g_{\alpha \beta} (r)$ become more pronounced. The height of the
first-nearest-neighbor peak changes by about 20 \%! The amount of this
change is significantly larger than any change observed for
macroscopic properties~[6] and thus we conclude that microscopic properties
can show a much stronger dependence on the cooling rate than the
macroscopic properties do.

From Fig. 2 one also sees that although the {\it height} of the
various peaks shows a significant cooling-rate dependence, the {\it
location} of the peaks is affected much less by the variation of
$\gamma$. Thus it is reasonable to compare these positions with
experimental data, indicated in broken lines. E.g., Mozzi and Warren
[21] quoted for the location of the $1^{\mbox{\scriptsize st}}$ and
$2^{\mbox{\scriptsize nd}}$ peak 1.62, 4.15; 2.65, 4.95; 3.12, 5.18
{\AA} for SiO, OO and SiSi, respectively. Although the agreement is not
perfect, the BKS potential does quite well to reproduce the short and
medium-range order of silica glass. Note that this potential was
developed to describe the {\it crystalline} modifications of SiO$_2$,
and in Ref. [6] it was used for amorphous SiO$_2$ without adjusting
any parameter whatsoever.

Similar conclusions emerge from a discussion of the partial structure
factors $S_{\alpha \beta}(q)$ [6], where $q$ is the scattering 
wavenumber, which is the structural quantity that is most directly
accessible in diffraction experiments. But the advantages of the
simulation is that full microscopic details of the structure are much
more easily accessible.  Since Fig. 2 shows that the location
$r_{\mbox{\scriptsize min}}$ of the first minimum in the radial
distribution function is practically independent of $\gamma$ (namely
$r^{{\rm SiSi}}_{\mbox{\scriptsize min}} = 3.42$~{\AA},
$r^{{\rm SiO}}_{{\mbox{\scriptsize min}}} = 2.20$~{\AA}, and
$r^{{\rm OO}}_{\mbox{\scriptsize min}} = 3.00$~{\AA}), one can compute
unambiguously the (partial) coordination number $z$ of particle $i$ by
defining $z$ as the number of other particles $j$ with $|\vec{r}_j -
\vec{r}_i| < r_{\mbox{\scriptsize min}}$. Fig. 3 shows
$P^{z=n}_{\alpha \beta}$, the probability that a particle of type
$\alpha$ has $n$ nearest neighbors of type $\beta$.  The vast majority
of silicon atoms are surrounded by $z=4$ oxygens; the case $z=5$ occurs
for the fast cooling rate with probability 5 \%  and decreases to about
0.5 \% probability for the slowest cooling rate.  Similarly $z=2$
dominates for oxygens. Thus the BKS potential automatically yields the
``rules'' commonly postulated for ideal amorphous SiO$_2$, namely that
this system is a ``continuous random network'' with these coordination
numbers $z=4$ for Si and $z=2$ for O~[8].

More disorder, however, is found on an intermediate length scale: while
each SiO$_4$ tetrahedron is surrounded by four other tetrahedra on
average, even for the smallest cooling rate the probability
$P^{z=5}_{{\rm SiSi}}$ is still about $1.5 \%$, and $P^{z=6}_{{\rm
OO}}$ increases only from about $0.6$ to $0.87$ with decreasing cooling
rate, over the accessible range of $\gamma$ (about $10 \%$ of the
oxygen atoms have seven other oxygen atoms as second nearest neighbors
instead of six).  The spread in these next-nearest neighbor
coordinations in the network is a clear indication of disorder.

This disorder also shows up clearly in the distribution functions of
various bond angles (Fig. 4). It is remarkable that the distribution
$P_{{\rm OSiO}} (\theta)$ has its maximum at an angle close to the
value of an ideal tetrahedron (109.47$^{\circ}$), and that the width of
this distribution clearly narrows with decreasing cooling rate. As far
as experimental data on the peak positions of these angular
distributions and their halfwidths are available at all, the
simulations for the slowest cooling rate are already very close to the
experimental values. This shows that such simulations can give quite
reliable information on such quantities.

Another characteristic of the medium range order in glasses, that is
much discussed in the literature, is the distribution of the frequency
of {\it rings} of a given size. A ring is defined as follows: starting
from a Si atom one chooses two different O atoms that are nearest
neighbors. Pick one of these. In general this O atom will also be a
nearest neighbor of a second Si atom. From this new Si atom one then
picks a new nearest-neighbor O atom, etc. This process is continued
until one returns to the O atom which is the second one of the
nearest-neighbor O atoms of the first Si-atom. In this way one has
constructed a closed loop of Si--O segments. The shortest one of these
loops is called the ring associated with the original Si atom and the
two nearest-neighbor O-atoms. The number of Si--O segments in this loop
is called the size $n$ of this ring.

Fig. 5 shows $P (n)$, the probability that a ring has length $n$, as
function of the cooling rate.  As expected, the case $n=6$ dominates,
and $P(n)$ for small $n$ $(n=3,4)$ clearly decrease with decreasing
$\gamma$. Note that in $\beta$-cristobalite (the first crystalline
phase reached from the liquid) only $n=6$ occurs, while the
lower-temperature phases $\beta$-tridymite and $\beta$-quartz have
rings with both $n=6$ and $n=8$, but no odd-numbered rings. The large
numbers of rings with $n=5$ and $n=7$ can be considered as a hallmark
of amorphous silica, and the difficulty to eliminate these ``defects''
from the structure prevents the system to crystallize easily.

\subsubsection*{Concluding remarks}   
In the previous subsection, it was demonstrated (choosing the work
[6] as an example) that with the slowest cooling rates just
available, it is now possible to create ``good'' amorphous silica
structures, which agree surprisingly well with experimental findings,
despite the fact that the simulated cooling rate exceeds the
experimental one by many orders of magnitude. However, clear
discrepancies remain when one considers the temperature dependence of
various quantities, because first of all the system falls out of
equilibrium at a much higher temperature than in the experiment, and
secondly because quantum effects are missing and they are expected
to become increasingly important at low temperatures. Of course, we
expect that, if the length scales we consider increase, we will find a
stronger dependence of the medium range order on the cooling rates,
some evidence in this direction was already discussed here, and the
more the simulated model will differ from reality (since quantitative
information on properties like the ring statistics, Fig. 5, is not
available from experiment, this expected discrepancy is not yet
detectable in practice: pair correlations $g_{\alpha \beta} (r)$ or
structure factors $S_{\alpha \beta} (q)$ simply are too insensitive
probes of amorphous structures!).

Nevertheless, the rich information from the MD simulation on a large
variety of structural properties \{$g_{\alpha \beta}(r)$, partial
coordination numbers, bond angle distributions, ring statistics,
etc.\}, and the dependence of these properties on both temperature and
cooling rate, allows to correlate these properties with each other and
thus to gain insight into how and why the amorphous structure forms. We
feel that this insight still may be partially unraveled, and additional
studies are desirable (including studies with other potentials that
have been used in the literature).

One major challenge, of course, is the extension of these studies from
pure SiO$_2$ to silicate glasses that contain more components. One
obvious obstacle to simulate, say, a mixture of SiO$_2$ and B$_2$O$_3$ is
that presently the interaction of a pair of oxygen in SiO$_2$ is modelled by
a completely different function [12] than in B$_2$O$_3$ [10,11].
But promising first steps to  model more-component glasses such as
sodium silicate glasses have been taken a long time ago [22], and
we think time is ripe to resume such efforts.

 \section*{Acknowledgements:}
This brief review is largely based on research done jointly with K.
Vollmayr \{Ref.[6]\}. It is a great pleasure to thank her for a very
pleasant and fruitful collaboration. We also have significantly
profited from collaborations with H.C. Andersen, C.A. Angell, J.
Horbach, J.L. Barrat, and others. Stimulating and informative
discussions with U. Buchenau, U. Fotheringham, D. Krause, W. Pannhorst
and R. Sprengard also deserve mention. Last but not least we thank the
SCHOTT Glaswerke Fonds and the Deutsche Forschungsgemeinschaft (DFG,
grant No. SFB 262/D1) and the Bundesministerium f\"ur Bildung,
Forschung, Wissenschaft und Technologie (BMBF, grant No. 03N8008C) for
financial support for this research.

\newpage

\section*{References}
\begin{description}
\item{1}
W. Kohn: ``Density functional theory'', in {\it Monte Carlo and
             Molecular Dynamics of Condensed Matter Systems}, ed. by
             K. Binder, G. Ciccotti (Societ{\'{a}} Italiana di Fisica,
             Bologna, 1996) pp. 561-572
\item{2}
   R. Car: ``Molecular dynamics from first principles'', in {\it Monte Carlo
             and Molecular Dynamics of Condensed Matter Systems}, ed. by
             K. Binder, G. Ciccotti (Societ{\'{a}} Italiana di Fisica,
             Bologna, 1996) pp. 601-634
\item{3}
   P. Nielaba: ``Quantum simulations in materials science: Molecular
             monolayers and crystals'', in {\it Annual Reviews of
             Computational Physics V}, ed. by D. Stauffer (World Scientific,
             Singapore, 1997) pp. 137-199
\item{4}
   M. Sprik: ``Introduction to molecular dynamics methods'', in
             {\it Monte Carlo and Molecular Dynamics of Condensed Matter
             Systems}, ed. by K. Binder, G. Ciccotti (Societ{\'{a}} Italiana
             di Fisica, Bologna, 1996) pp. 43-88
\item{5}
   C. A. Angell, J. H. R. Clarke, L. V. Woodcock: ``Interaction potentials
             and glass formation: a survey of computer experiments'', in
             {\it Advances in Chemical Physics}, Vol. XLVIII, ed. by
             I. Prigogine, S. Rice (J. Wiley, New York, 1981) pp. 397-453
\item{6}
   K. Vollmayr, W. Kob, K. Binder: ``Cooling-rate effects in amorphous
             silica: A computer simulation study'' Phys. Rev. B {\bf 54},
             15808-15827 (1996)
\item{7}
   W. Kob: ``Computer Simulations of Supercooled Liquids and Structural
             Glasses'', in {\it Annual Reviews of Computational Physics III},
             ed. by D. Stauffer (World Scientific, Singapore, 1995), pp. 1-43
\item{8}
   R. Zallen: ``Stochastic geometry: Aspects of amorphous solids'', in
             {\it Fluctuation Phenomena}, ed. by E. W. Montroll, J. L. Lebowitz
             (North-Holland, Amsterdam, 1979), pp. 177-228
\item{9}
   K. Binder: ``Applications of Monte Carlo Methods to Statistical Physics''
              Rep. Progr. Phys. {\bf 60}, 487-559 (1997)
\item{10}
   A. Takada, C. R. A. Catlow, G. D. Price: ``Computer modelling of
              B$_2$O$_3$: part I. New interatomic potentials, crystalline
              phases, and predicted polymorphs'' J. Phys.: Cond. Matter
              {\bf 7}, 8659-8692 (1995)
\item{11}
   A. Takada, C. A. R. Catlow, G. D. Price: ``Computer modelling of
              B$_2$O$_3$: part II. Molecular dynamics simulations of
              vitreous structures'' J. Phys.: Cond. Matter {\bf 7},
              8693-8722 (1995)
\item{12}
   B. W. H. van Beest, G. J. Kramer, R. A. van Santen: ``Force fields for
              silicas and aluminophosphates based on {\it ab initio}
              calculations'', Phys. Rev. Lett. {\bf 64}, 1955-1958 (1990)
\item{13}
   J. Horbach, W. Kob, K. Binder: ``Molecular dynamics simulations of the
              dynamics of supercooled silica'', Phil. Mag. B (in press)
\item{14}
   J. Horbach, W. Kob, K. Binder: ``The dynamics of supercooled silica:
              acoustic modes and boson peak'', J. Non-Cryst. Solids
              (in press)
\item{15}
   J. Horbach, W. Kob, K. Binder, C. A. Angell: ``Finite size effects in
              simulations of glass dynamics'', 
              Phys. Rev. E {\bf 54}, R5897-R5900 (1996)
\item{16}
   U. Buchenau: ``Neutron scattering at the glass transition'', in
              {\it Phase Transitions and Relaxation in Systems with
              Competing Energy Scales}, ed. by T. Riste and D. Sherrington
              (Kluwer Acad. Publ., Dordrecht, 1993) pp. 233-257
\item{17}
   O. V. Mazurin, M. V. Streltsina, T. P. Shvaikoskaya: {\it Handbook of
              Glass Data} (Elsevier, Amsterdam, 1983) Part A
\item{18}
   B. Vessal, M. Amini, D. Fincham, C. R. A. Catlow: ``Water-like melting
              behavior of SiO$_2$ investigated by the molecular dynamics
              simulation technique'', Phil. Mag. B {\bf 60}, 753-775 (1989)
\item{19}
   R. Br\"uckner: ``Properties and Structure of Vitreous Silica. I'' 
              J. Non-Cryst. Solids {\bf 5}, 123-175 (1970)
\item{20}
   J.-P. Hansen, I.R. McDonald: {\it Theory of Simple Liquids} (Academic,
              London, 1986)
  \item{21}
   R. L. Mozzi, B. E. Warren: ``The Structure of Vitreous Silica''
              J. Appl. Cryst. {\bf 2}, 164-172 (1969)
\item{22}
   T. F. Soules ``A molecular dynamics calculation of the structure of
              sodium silicate glasses'', J. Chem.  Phys. {\bf 71},
              4570-4578 (1979)

 \end{description}

\newpage

\begin{figure}[f]
\psfig{file=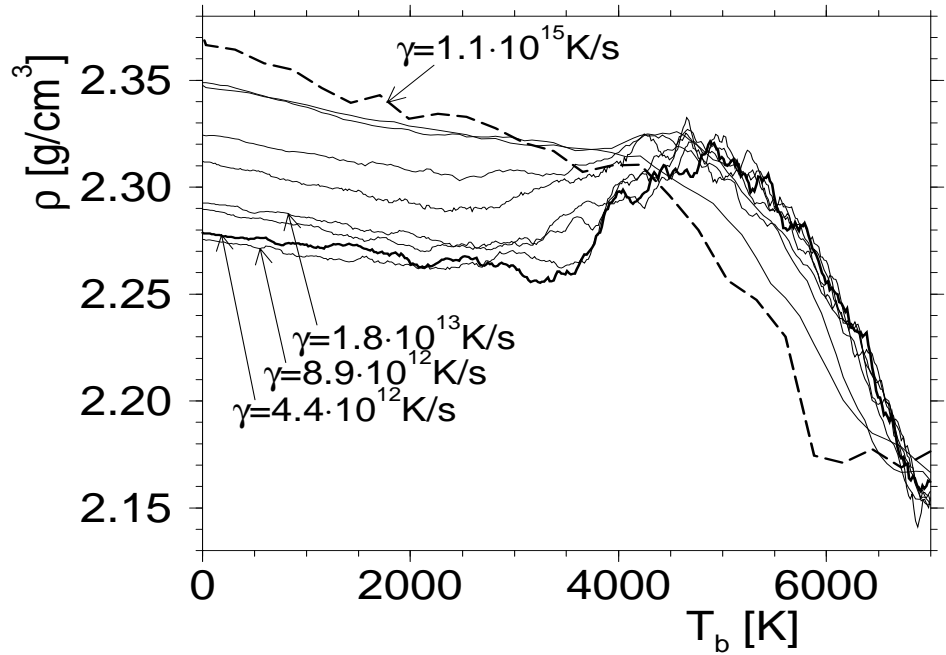,width=13cm,height=9.5cm}
\caption{
Density of simulated SiO$_2$ vs. bath temperature $T_b$, for all
cooling rates investigated. The solid and dashed bold curves are the
smallest and largest cooling rates $\gamma$, respectively. Note the
presence of a local maximum in $\rho$ at temperatures around 4800 K
if $\gamma$ is small. From {\it Vollmayr et al.}~[6].}
\end{figure}
\begin{figure}[f]
\psfig{file=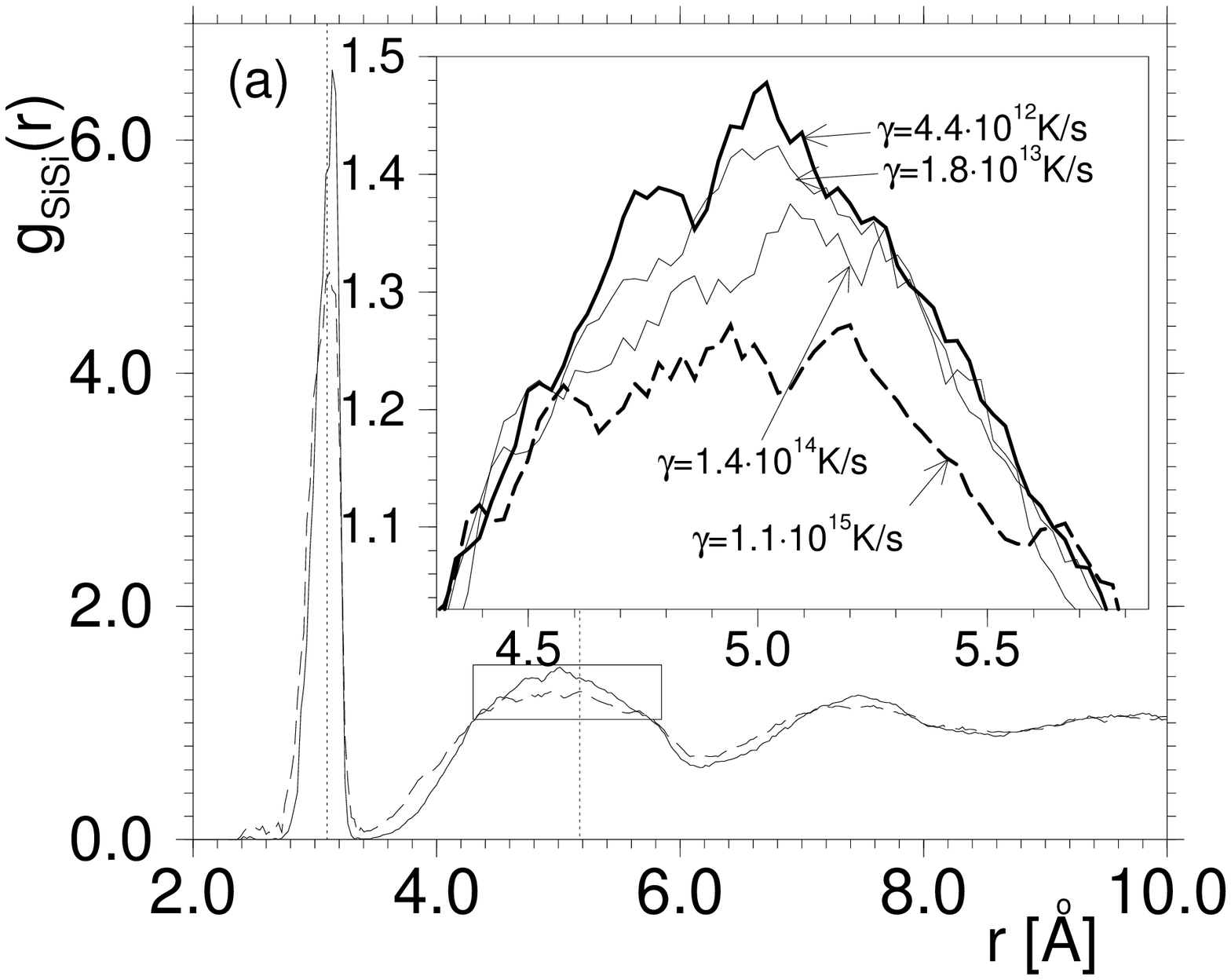,width=13cm,height=9.5cm}
\end{figure}
\begin{figure}[f]
\psfig{file=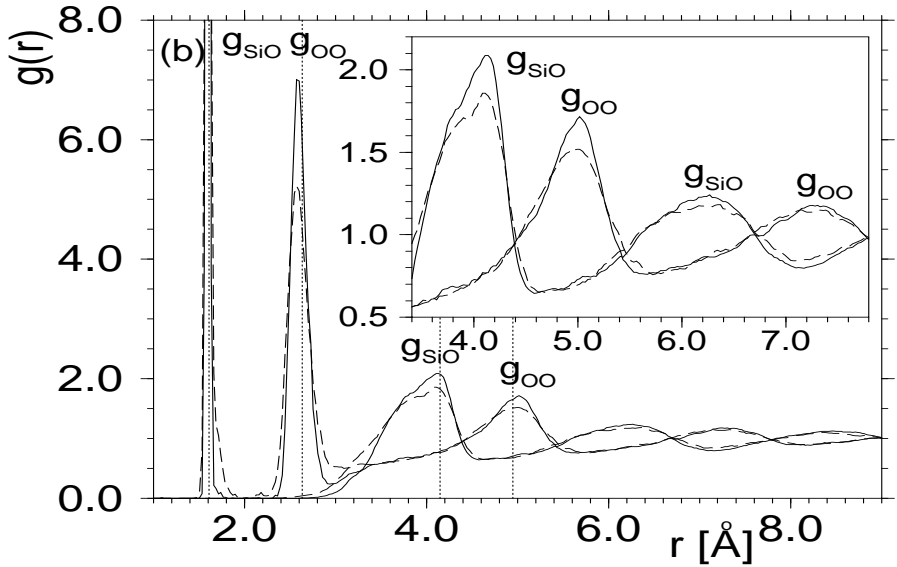,width=13cm,height=8.5cm}
\caption{
Radial distribution functions of simulated amorphous SiO$_2$. (a)
$g_{{\rm SiSi}}(r)$. Main figure: the slowest (solid curve) and fastest
(dashed curve) cooling rate. The vertical dotted lines give the
position of the peak as determined from experiments (see text).
Inset:  enlargement of the second-nearest neighbor peak for four
selected cooling rates. (b) $g_{{\rm SiO}}(r)$ and $g_{{\rm OO}}(r)$
for the slowest (solid curves) and fastest (dashed curves) cooling
rate.  Inset: enlargement of the second-and third-nearest-neighbor
peak. From {\it Vollmayr et al.}~[6].}
\end{figure}
\begin{figure}[f]
\psfig{file=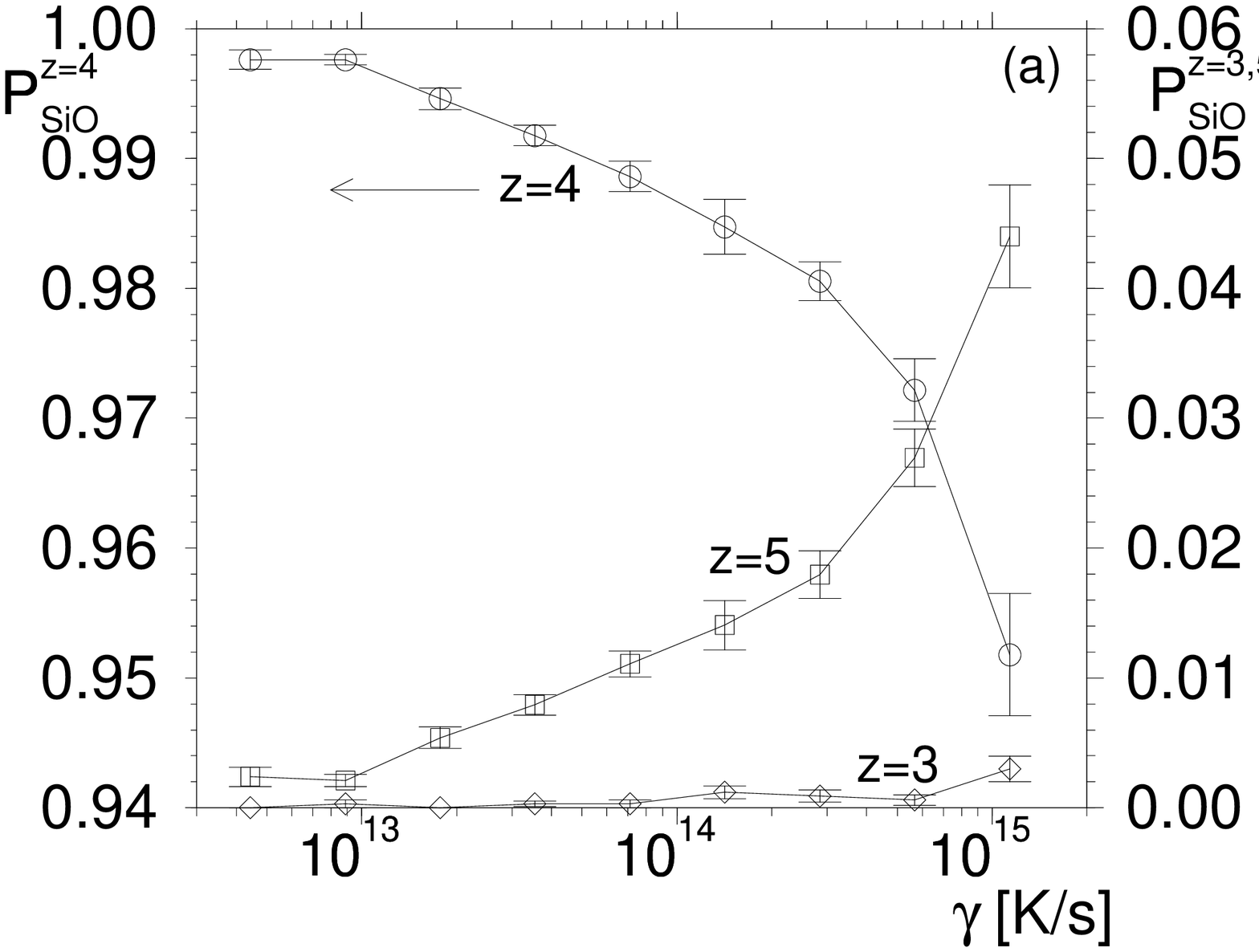,width=13cm,height=8.5cm}
\end{figure}
\begin{figure}[f]
\psfig{file=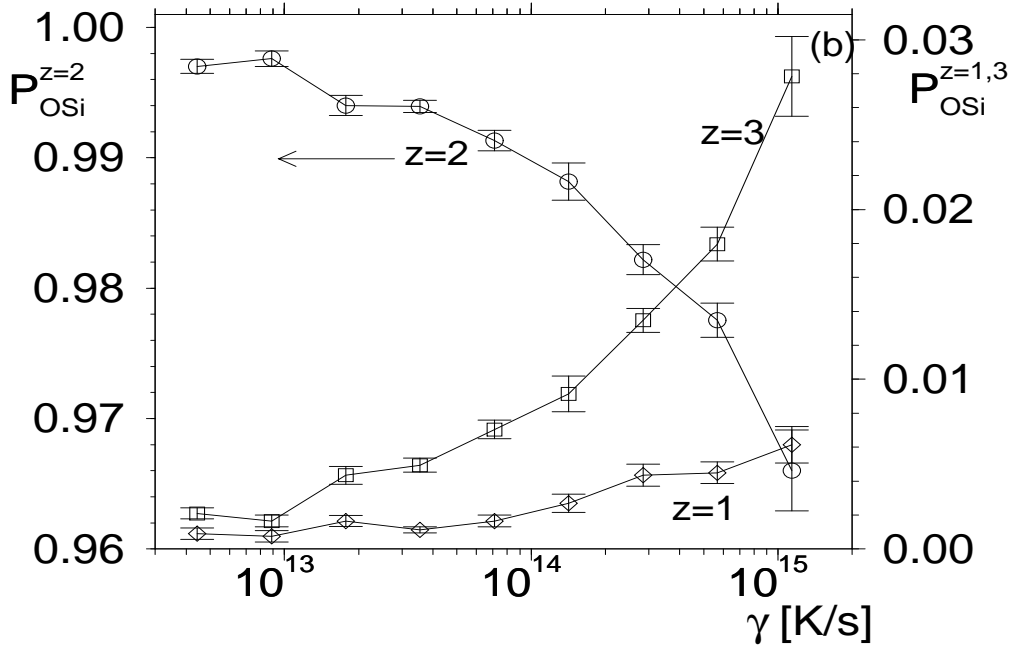,width=13cm,height=9.5cm}
\caption{
Probabilities of partial coordination numbers $P^{z=n}_{\alpha \beta}$
plotted vs. cooling rate, for Si--O pairs (a) and O--Si pairs (b).
Note the different scales for the various curves. From {\it Vollmayr
et al.}~[6].}
\end{figure}
\begin{figure}[f]
\psfig{file=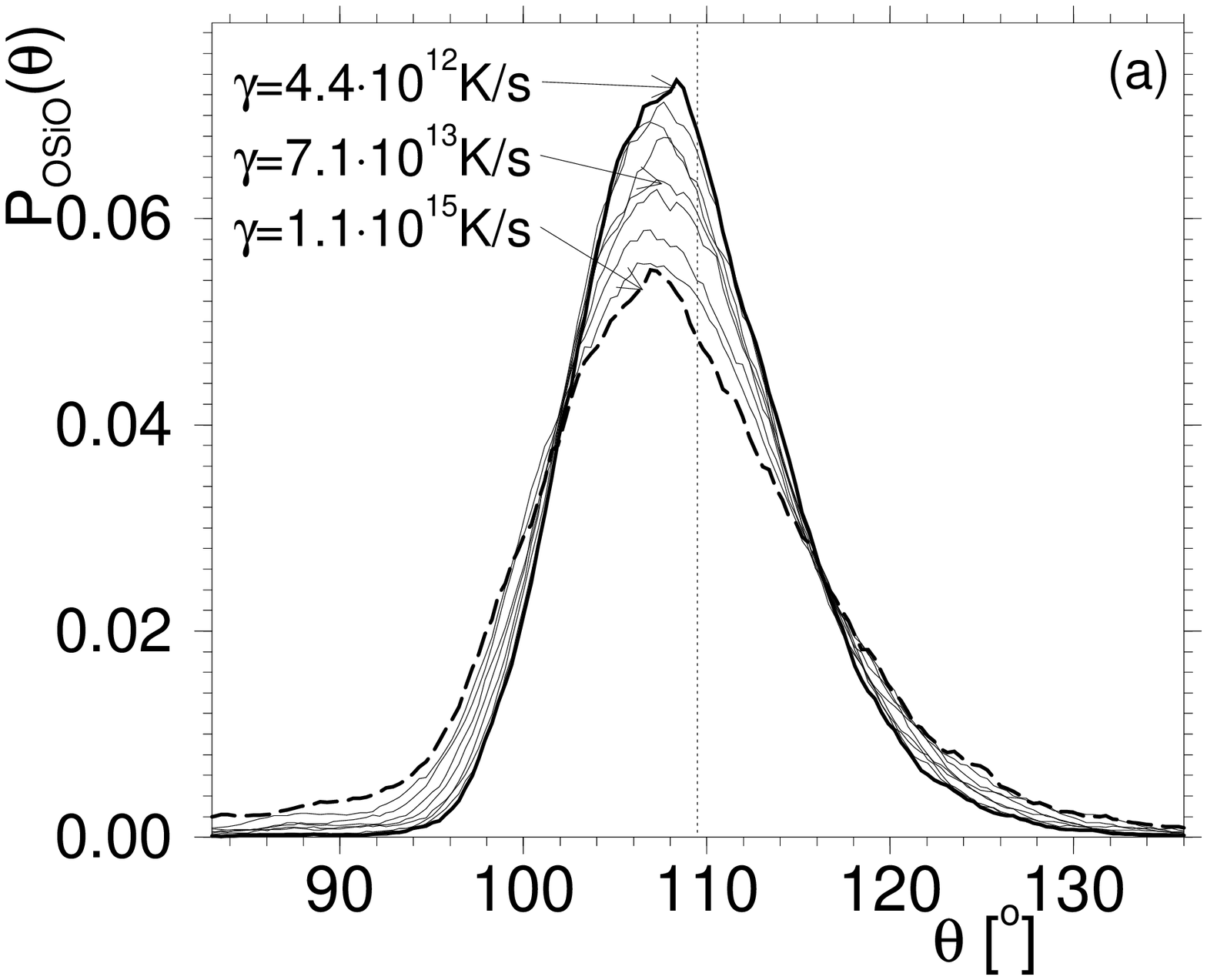,width=13cm,height=9.5cm}
\end{figure}
\begin{figure}[f]
\psfig{file=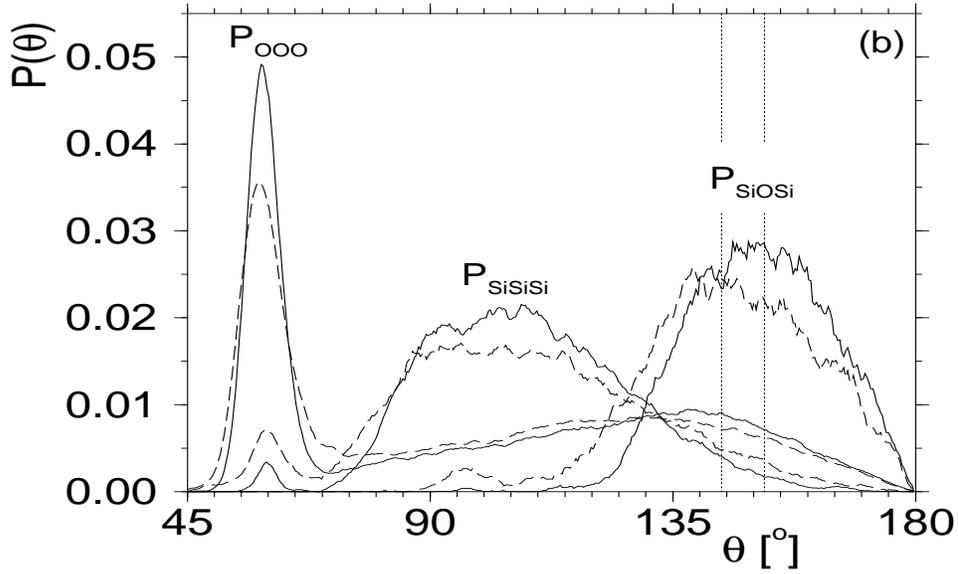,width=13cm,height=8.5cm}
\caption{
Distribution function of various angles at several cooling rates in
amorphous SiO$_2$ at $T=0$. (a) Angle O-Si-O for all cooling rates
studied. The vertical line is the experimental value [21].
(b) Angles O-O-O, Si-Si-Si and Si-O-Si for the slowest
(solid curves) and fastest (dashed curves) cooling rates investigated.
Vertical lines are experimental values from different authors
(see [6] for details.). From {\it Vollmayr et al.}~[6].}
\end{figure}
\begin{figure}[f]
\psfig{file=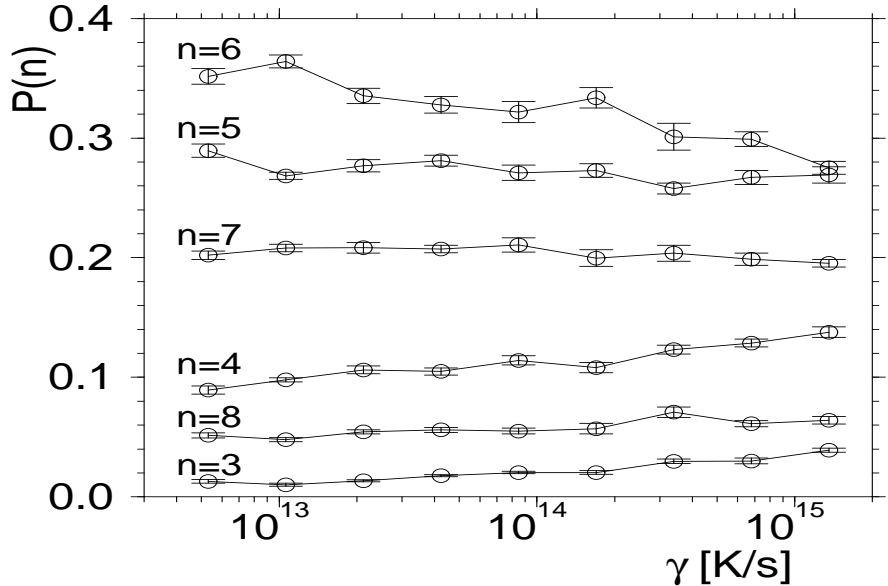,width=13cm,height=8.5cm}
\caption{
Probability $P(n)$ that a particle is a member of a ring of size
$n$, plotted vs. $\gamma$. From {\it Vollmayr et al.}~[6].}
\end{figure}
\newpage

 \end{document}